\documentclass[aps,pre,preprint]{revtex4}
\usepackage{graphicx}
\usepackage{amsfonts}
\usepackage{amssymb}
\usepackage{amsmath}

\begin{document}

\title{\bf Matter-wave solitons of collisionally inhomogeneous condensates}

\author{
G.\ Theocharis$^{1,4}$,
P.\ Schmelcher$^{1,2}$, 
P.G.\ Kevrekidis$^{3}$ and 
D.J.\ Frantzeskakis$^{4}$ 
}
\affiliation{
$^{1}$ Theoretische Chemie, Physikalisch-Chemisches Institut, Im Neuenheimer Feld 229,
Universit\"at Heidelberg, 69120 Heidelberg, Germany \\
$^{2}$ Physikalisches Institut, Philosophenweg 12, Universit\"at Heidelberg, 69120 Heidelberg, Germany \\
$^{3}$ Department of Mathematics and Statistics,University of Massachusetts, Amherst MA 01003-4515, USA \\
$^{4}$ Department of Physics, University of Athens, Panepistimiopolis,Zografos, Athens 157 84, Greece }

\begin{abstract}
We investigate the dynamics of matter-wave solitons in the presence of a spatially varying
atomic scattering length and nonlinearity. The dynamics of bright and dark 
solitary waves 
is studied 
using the corresponding Gross-Pitaevskii equation. 
The numerical results are shown 
to be in very good agreement with the predictions of the 
effective equations of motion derived by
adiabatic perturbation theory. The spatially dependent 
nonlinearity leads to a gravitational potential
that allows to influence the motion of both fundamental 
as well as higher order solitons. 
\end{abstract}

\maketitle 


\section{Introduction}

Recent years have seen enormous progress with respect to our understanding and the controlled
processing of atomic Bose-Einstein condensates (BECs) \cite{dalfovo} both for theory and experiment.
In case of nonlinear excitations, specifically solitons, the 
experimental observation of dark \cite{dark}, bright \cite{expb1,expb2} and gap \cite{gap} solitons 
has inspired many studies on matter-wave solitons in general. 
Apart from a fundamental interest in their behavior and properties, solitons are potential candidates
for applications since there are possibilities to coherently manipulate them in matter-wave devices, such as 
atom chips \cite{Folman}. Moreover, the formal similarities between matter-wave and optical solitons indicate that 
the former may be used in future applications similarly to their optical siblings, which have a time-honored history 
in optical fibers and waveguides (see, e.g., the recent reviews \cite{buryak,kivpr}). 

Typically dark (bright) matter-wave solitons are formed in atomic condensates with repulsive (attractive) 
interatomic interactions, i.e. for atomic species with positive (negative) scattering length $a$. 
One of the very interesting aspects for tailoring and designing the properties of (atomic or molecular) BECs 
is the possibility to control the interaction of ground state species by changing the threshold collision
dynamics and consequently changing either the sign or the magnitude of the scattering length. A prominent
way to achieve this is to apply an external magnetic field which provides control over the scattering
length because of the rapid variation in collision properties associated with a threshold scattering resonance
being a Feshbach resonance (see refs. \cite{Weiner03,feshbach} and references therein).
For low-dimensional setups a complementary way of tuning the scattering length or 
the nonlinear coupling at will is to change the transversal confinement in order to achieve an effective
nonlinearity parameter for the dynamics in e.g. the axial direction. In the limit of very strong transversal
confinement this leads to the so-called confinement induced resonance at which the modified scattering
length diverges \cite{Olshanii98}. A third alternative approach uses the possibility of tuning the
scattering length with an optically induced Feshbach resonance \cite{Theis04}. Varying the interactions
and collisional properties of the atoms was crucial for a variety of experimental discoveries such as 
the formation of molecular BECs \cite{Herbig03} or the revelation of the BEC-BCS crossover \cite{Bartenstein04}.
Recent theoretical studies have predicted that a time-dependent modulation of the 
scattering length can be used to prevent collapse in higher-dimensional attractive BECs \cite{FRM1}, or to create robust 
matter-wave solitons \cite{FRM2}. 

Reflecting the increasing degree of control with respect to the processing of BECs it is nowadays ``not only''
possible to change the scattering length in the same way for the complete ultracold atomic ensemble i.e. it can be
tuned globally, but it is possible to obtain a locally varying scattering length thereby providing
a variation of the collisional dynamics across the condensate. According to the above this can be implemented by a
(longitudinally) changing transversal confinement or an inhomogeneity of the external magnetic field
in the vicinity of a Feshbach resonance. There exist only very few investigations on condensates 
in such an inhomogeneous environment \cite{fka,fermi}.

To substantiate the above, let us specify in some more detail the case of a magnetically tuned scattering length.
The behavior of the scattering length near a Feshbach resonant magnetic field $B_0$ 
is typically of the form $a(B)=\tilde{a}[1-\Delta/(B-B_{0})]$, where 
$\tilde{a}$ is the value of the scattering length far from resonance and $\Delta$ represents the width of the 
resonance (see, e.g., \cite{fa}). Let us consider a quasi-1D condensate along the $x$-direction exposed to
a bias field $B_1$ sufficiently far from the resonant value $B_0$ in the presence of an additional gradient $\epsilon$
of the field i.e. we have $B=B_{1}+\epsilon x$ such that $B_{1}>B_{0}+\Delta$ (without loss of generality we
take $\epsilon > 0$). Assuming $\epsilon x/(B_{1}-B_{0}) \ll 1$ for all values of $x$ in the interval $(-L/2, L/2)$, 
where $L$ is the characteristic spatial scale on which the evolution of the condensate takes place, 
it is readily seen that the scattering length can be well approximated by the spatially dependent form 
$a(x)=a_{0}+a_{1}x$, where $a_{0}=\tilde{a}[1-\Delta/(B_{1}-B_{0})]$ and $a_{1}= \epsilon \Delta \tilde{a}/(B_{1}-B_{0})^{2}$. 
In the following we will assume that $a_0$ and $a_1$ are of the same sign.

This opens the perspective of studying collisionally inhomogeneous condensates. 
In this work, we provide a first step in this direction by investigating the behavior of
nonlinear excitations, specifically bright and dark matter-wave solitons 
in attractive and repulsive quasi one-dimensional (1D) BECs, in the presence of a spatially-dependent
scattering length and nonlinearity. We investigate the soliton dynamics in different setups
and analyze the impact of the spatially varying nonlinearity by numerically integrating
the Gross-Pitaevskii (GP) equation as well as in the framework of adiabatic perturbation theory for solitons \cite{kima,KY}.

The paper is organized as follows: In Sec. II the effective perturbed NLS equation is derived. In Sec. III, fundamental 
and higher-order soliton dynamics are considered and Bloch oscillations in the additional presence of an optical lattice are studied. 
Sec. IV is devoted to the study of dark matter-wave solitons, and in Sec. V the main findings of this work are summarized.

\section{The perturbed NLS equation}

At sufficiently low temperatures, the dynamics of a quasi-one-dimensional BEC aligned along
the $x$--axis, is described by an 
effective one-dimensional (1D) GP equation (see, e.g., \cite{gp1d}) of the form:
\begin{equation}
i \hbar \frac{\partial \psi}{\partial t} = - \frac{\hbar^{2}}{2m} \frac{\partial^{2} \psi}{\partial x^{2}} 
+ V(x)\psi + g |\psi|^{2} \psi,  
\label{dimgpe}
\end{equation}
where $\psi(x,t)$ is the order parameter, $m$ is the atomic mass, and $V(x)$ is the external potential.
Here we assume that the condensate is confined in a harmonic trap i.e., 
$V(x) =(1/2) m \omega_{x}^{2} x^{2}$
where $\omega_{x}$ is the confining frequency in the axial direction.
The nonlinearity coefficient $g$, accounting for the interatomic interactions, has an effective 1D form, namely 
$g=2 \hbar a \omega_{\perp}$, where $\omega_{\perp}$ is the transverse-confinement frequency and $a$ is the atomic
s-wave scattering length. The latter is positive (negative) for repulsive (attractive)
condensates consisting of e.g. $^{87}$Rb ($^{7}$Li) atoms.

As discussed in the introduction we assume a collisionally inhomogeneous 
condensate i.e., a spatially
varying scattering length according to $a(x)=a_{0}+a_{1}x$ where $a_{0}$ and $a_1$ are both positive (negative) for 
repulsive (attractive) condensates. Moreover, if the characteristic length $L$ for the evolution 
of the condensate implies $|a_{1} L| \le |a_{0}|$, it is readily seen that the function $a(x)$ can be expressed as 
$a(x)=s A(x)$, where $A(x) \equiv |a_{0}|+|a_{1}|x $  is a positive definite function (for $-L/2 < x < L/2$)
and $s=\rm{sign} (a_{0})= \pm 1$ for repulsive and attractive condensates respectively. We can then 
reduce the original GP Eq. (\ref{dimgpe}) to a dimensionless form as follows: 
$x$ is scaled in units of the healing length $\xi=\hbar/\sqrt{n_{0} g_{0} m}$, $t$ in units of $\xi/c$ 
(where $c=\sqrt{n_{0}g_{0}/m}$ is the Bogoliubov speed of sound), 
the atomic density 
$n \equiv |\psi|^{2}$ is rescaled by the peak density $n_{0}$, 
and energy is measured in units of the chemical potential of the system $\mu=g_{0} n_{0}$; in the above expressions 
$g_{0} \equiv 2 \hbar a_{0} \omega_{\perp}$ corresponds to the constant (dc) value $a_{0}$ of the scattering length. 
This way, the following normalized GP equation is readily obtained, 
\begin{equation}
\label{GP} 
i \frac{\partial \psi}{\partial t}= - \frac{1}{2} \frac{\partial^{2} \psi }{\partial x^{2}} 
+ V(x) \psi + s g(x)|\psi|^2 \psi, 
\end{equation}
where $V(x)=(1/2) \Omega^{2} x^{2}$ and the parameter $\Omega \equiv (2 a_0 n_{0})^{-1}(\omega_{x}/\omega_{\perp})$ 
determines the magnetic trap strength. Additionally, $g(x)=1+\delta x$ is a positive definite function and 
$\delta \equiv \epsilon \Delta (B_{1}-B_{0})^{-1} \left[1-\Delta (B_{1}-B_{0})\right]^{-1}$ is the gradient.

Let us assume typical experimental parameters for a quasi-1D condensate containing $N \sim 10^{3}$ 
atoms and with a peak atomic density $n_{0} \approx 10^8$ m$^{-1}$. Then, taking the scattering length $a$ to be of the
order of a nanometer we assume that the 
ratio of the confining frequencies $\omega_{x}/\omega_{\perp}$ varies between 0.01 and 0.1.
Therefore, the trap 
strength $\Omega$ is typically O$(10^{-2})$-O$(10^{-1})$. Furthermore, we 
will assume that the field gradient $\epsilon$ 
is also small and accounts for the leading order corrections of the gradient $\delta$
in what follows.
Thus, $\Omega$ and $\delta$ are the natural small parameters 
of the problem. 

We now introduce the transformation 
$\psi=u / \sqrt{g}$ to rewrite Eq. (\ref{GP}) in the following form:  
\begin{eqnarray} 
i \frac{\partial u}{\partial t}+\frac{1}{2} \frac{\partial^{2} u}{\partial x^{2}} - s |u|^{2} u = R(u).
\label{gpe1d_u} 
\end{eqnarray} 
Apparently, Eq. (\ref{gpe1d_u}) has the form of a perturbed NLS equation
(of the focusing or defocusing type, for $s=-1$ and $s=+1$ respectively), with the perturbation $R(u)$ being given by  
\begin{eqnarray} 
R(u) &\equiv& V(x) u + \frac{d}{dx} \ln(\sqrt{g}) \frac{\partial u}{\partial x} \nonumber \\
&+& \frac{1}{2} \left[ \frac{d^2}{dx^2} \ln(\sqrt{g}) - \left(\frac{d}{dx} \ln(\sqrt{g})\right)^{2} \right]u.
\label{per}
\end{eqnarray} 


The last two terms on the right hand side of Eq. (\ref{per}) are of higher 
order with respect to the perturbation parameter $\delta$ than the second term 
and will henceforth be ignored (this will be discussed in more detail
below). We therefore examine the soliton 
dynamics in the presence of the perturbation including the first two terms of Eq. (\ref{per}). 

\section{Bright matter-wave solitons}

\subsection{Fundamental solitons}

In the case $s=-1$ and in the absence of the perturbation, Eq. (\ref{gpe1d_u}) 
represents the traditional 
focusing NLS equation, which 
possesses a commonly known family of fundamental bright soliton solutions
of the following form \cite{zsb},
\begin{eqnarray} 
u(x,t)=\eta {\rm sech}[\eta(x-x_{0})]\exp[i(kx-\phi(t)]
\label{ansatz} 
\end{eqnarray} 
where $\eta$ is the amplitude and inverse spatial width of the soliton, $x_{0}$ is the soliton center, 
the parameter 
$k=dx_{0}/dt$ defines both the soliton wavenumber and velocity, 
and finally $\phi(t)=(1/2)(k^2-\eta^2)t+\phi_{0}$ 
is the soliton phase ($\phi_0$ being an arbitrary constant). 
Let us assume now that the soliton width $\eta^{-1}$ is much smaller than $\Omega^{-1/2}$ and $\delta^{-1}$ [namely the characteristic 
spatial scales of the trapping potential and the function $g(x)$] or, physically speaking, the potential $V(x)$ and the function $g(x)$
vary little on the soliton scale. In this case, we may employ the 
adiabatic perturbation theory for solitons \cite{kima} 
to treat analytically the effect of the perturbation $R(u)$ 
on the soliton (\ref{ansatz}). According to this approach, the soliton 
parameters $\eta$, $k$ and $x_0$ become 
unknown, slowly-varying functions of time $t$, but the functional form of the soliton (see Eq. (\ref{ansatz})) remains unchanged. 
Then, from 
Eq. (\ref{gpe1d_u}), it is found that the 
number of atoms  
$N=\int_{-\infty}^{+\infty}|u|^{2}dx$ 
and the momentum 
$P=(i/2) \int_{-\infty}^{+\infty}\left[u (\partial u^{\star}/\partial x) - u^{\star} (\partial u/\partial x) \right] dx$
which are integrals of motion of the unperturbed system, evolve, in the presence of the perturbation, according to the following equations,
\begin{eqnarray} 
\frac{dN}{dt} &=& -2 {\rm Im} \left[ \int_{-\infty}^{+\infty} R u^{\star} dx \right], \,\,\, 
\label{evin1} \\
\frac{dP}{dt} &=& 2 {\rm Re} \left[ \int_{-\infty}^{+\infty} R \frac{\partial u^{\star}}{\partial x} dx \right].
\label{evin2}
\end{eqnarray}
We remark that the number of atoms $N$ is conserved for Eq. (\ref{GP}) but
the transformation, leading to Eq. (\ref{gpe1d_u}) no longer preserves
that conservation law, leading, in turn, to Eq. (\ref{evin1}).

We now substitute the ansatz (\ref{ansatz}) (but with the soliton parameters being functions of time) 
into Eqs. (\ref{evin1})-(\ref{evin2});
furthermore we use a Taylor expansion of the second term 
of Eq. (\ref{per}), around $x=x_0$ (keeping the two leading terms).
The latter expansion is warranted by the exponentional localization of
the wave around $x=x_0$. 
We then obtain the evolution equations for $\eta(t)$ and $k(t)$,
\begin{eqnarray} 
\frac{d \eta}{dt}&=&k \eta \frac{\partial}{\partial x_{0}} \ln(g), 
\label{bright_par1} \\
\frac{dk}{dt}&=&-\frac{\partial V}{\partial x_{0}} + \frac{\eta^2}{3} \frac{\partial }{\partial x_{0}}\ln(g).
\label{bright_par2} 
\end{eqnarray} 
To this end, recalling that $dx_{0}/dt=k$, we may combine Eqs. (\ref{bright_par1})-(\ref{bright_par2}) to derive 
the following equation of motion for the soliton center:
\begin{eqnarray} 
\frac{d^2x_{0}}{dt^2}=-\frac{\partial V}{\partial x_{0}}+\frac{\eta^{2}(0)}{6g^{2}(0)}\left(\frac{\partial g^2}{\partial x_{0}}\right), 
\label{eq_mot} 
\end{eqnarray} 
where $\eta(0)$ and $g(0) \equiv g(x_{0}(0))$ are the initial 
values of the amplitude and function $g(x)$ respectively. 
Notice that the above result indicates that the main contribution from 
the spatially dependent scattering length comes to order $\delta^2$ 
(while the contribution of the 
last two terms in Eq.
(\ref{per}) would have been
O$(\delta^3)$ and is neglected).
It is clear that in the particular case where 
$g(x)=1+\delta x$, Eq. (\ref{eq_mot}) describes the motion of a unit mass particle in the presence of the effective potential  
\begin{eqnarray} 
V_{\rm eff}(x_{0})=\frac{1}{2} \omega_{bs}^{2} x_{0}^{2}- \beta x_{0}, 
\label{Veff} 
\end{eqnarray}
where the parameter $\beta$ is defined as 
\begin{eqnarray}
\beta=\frac{\eta^{2}(0) \delta }{3\left[1+\delta x_{0}(0)\right]^2},
\label{lambda}
\end{eqnarray}
and  
\begin{equation}
\omega_{\rm bs}= \sqrt{\Omega^2-\delta \beta},
\label{bsf}
\end{equation}
is the oscillation frequency of the bright soliton. In the absence of the spatial variation of
the scattering length ($\delta=0$), Eq. (\ref{eq_mot}) 
actually expresses the Ehrenfest theorem, implying that the bright soliton oscillates with a frequency $\omega_{\rm bs}=\Omega$ 
in the presence of the harmonic potential with strength $\Omega$. Nevertheless, the presence of the gradient modifies 
significantly the bright soliton dynamics as follows: First, as seen by the second term in the right-hand side of Eq. (\ref{Veff}), 
apart from the harmonic trapping potential, an effective gravitational potential is also present, 
which induces an acceleration of the initial soliton towards larger values of $x_0$ (for $\delta >0$)
i.e. it shifts the center of the harmonic potential from $x_0=0$ to $x_0 = \frac{\beta}{\omega_{bs}^2}$.
Second, the oscillation frequency of the bright soliton 
is modified for $\delta \ne 0$, according to Eq. (\ref{bsf}). Moreover, depending on the initial values of the parameters, 
namely for $\beta \delta \ge \Omega^2$
an interesting situation may occur, in which the effective harmonic potential, instead of being purely attractive, 
it can effectively disappear, or be {\it expulsive}. 

The solution to Eq. (\ref{eq_mot}) in the variables $y_0=x_0-\beta/\omega_{bs}^2$ is, of course,
a simple classical oscillator 
\begin{eqnarray}
y_0(t)=y_0(0) \cos(\omega_{bs} t) + \frac{\dot{y}_0(0)}{\omega_{bs}} 
\sin(\omega_{bs} t)
\label{soln}
\end{eqnarray}
which is valid for $\omega_{bs}^2 >0$. For $\omega_{bs}^2 <0$ the trigonometric
functions have to be replaced by hyperbolic ones.
In the case $\omega_{bs}^2 =0$, the resulting motion (to the
order examined) is the one due to a uniform acceleration with
$x_0(t)=x_0(0) + \dot{x}_0(0) t + \beta t^2/2$.

The above analytical predictions have been confirmed by direct numerical simulations. 
In particular, we have systematically compared 
the results obtained from Eq. (\ref{eq_mot}) with the results of the direct numerical 
integration of the GP Eq. (\ref{GP}). In the following, we use the
trap strength $\Omega=0.05$, initial soliton amplitude $\eta(0)=1$ and initial location 
of the soliton $x_{0}(0)=0$, and different values for the 
normalized gradient $\delta$. The above 
values of the parameters, may correspond to a 
$^{7}$Li condensate containing 
$N \approx 4000$ atoms, confined in a quasi-1D trap 
with frequencies $\omega_{x}=2\pi \times 14$ Hz and 
$\omega_{\perp}=100 \omega_{x}$. Note that these values correspond to 
a scattering length $a=-0.21$ nm 
(pertaining to a magnetic field $425$ Gauss), a value for which a 
bright matter-wave soliton has been observed 
experimentally \cite{expb2}. 

\begin{figure}[tbp]
\includegraphics[width=6cm]{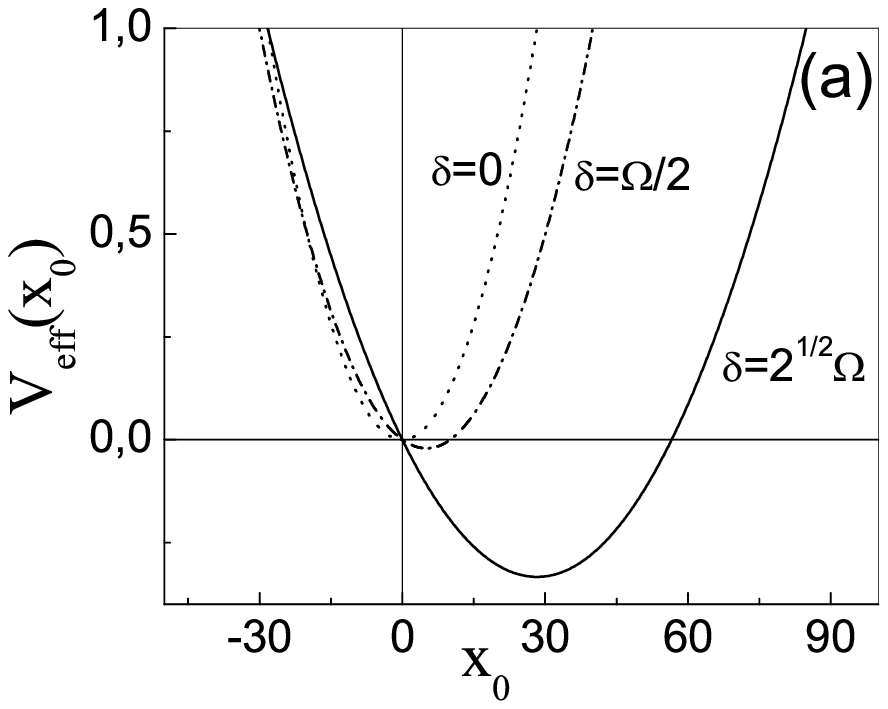}
\includegraphics[width=6cm]{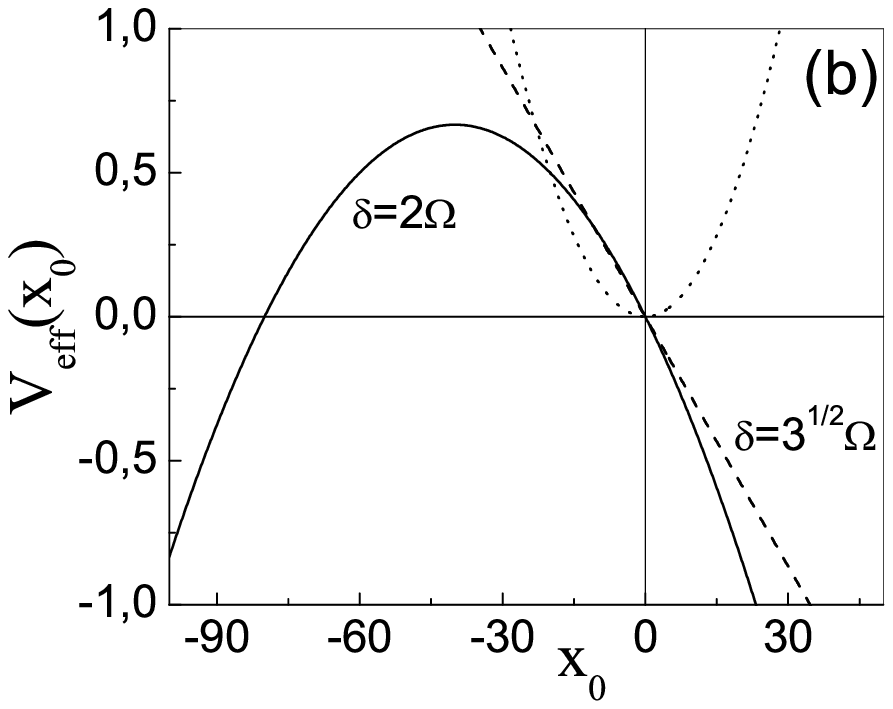}
\includegraphics[width=6cm]{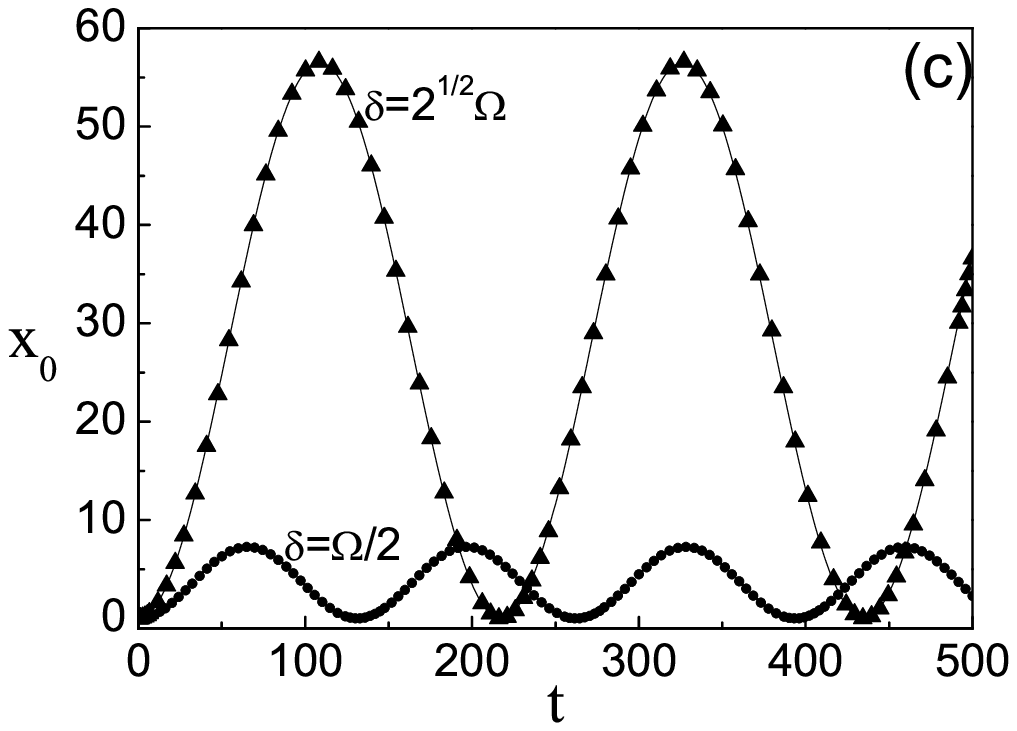}
\includegraphics[width=6cm]{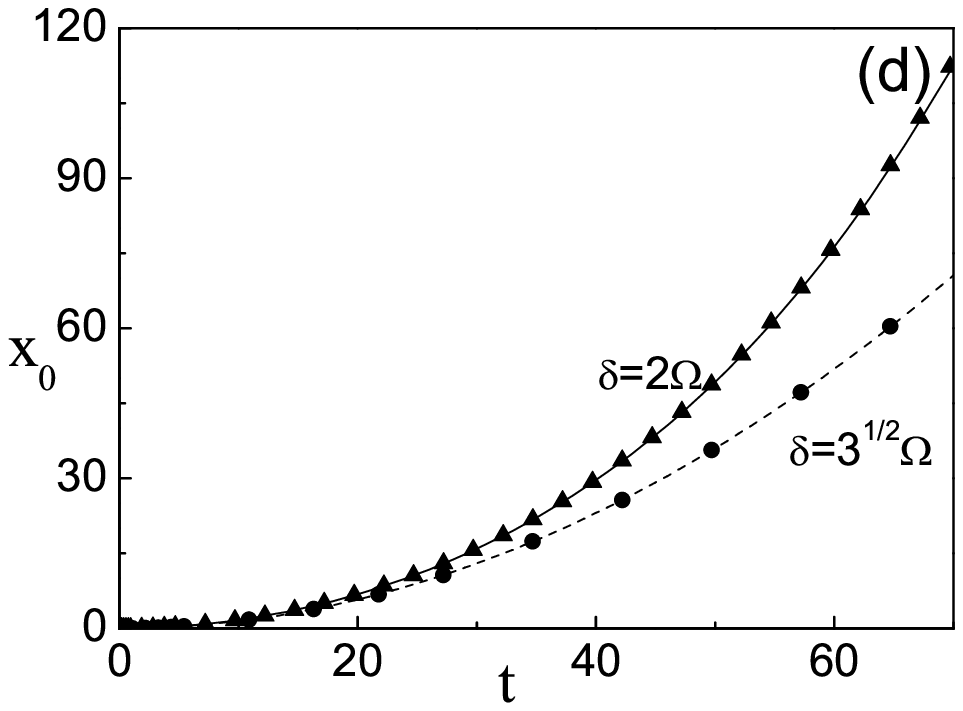}
\caption{Top panels: The effective potential $V(x_0)$ as a function 
of the soliton center $x_0$ 
for a trap strength $\Omega=0.05$ for a fundamental bright 
soliton of amplitude $\eta(0)=1$, 
initially placed at the trap center ($x_{0}(0)=0$). 
Different values of the gradient 
modify the character of the potential: 
in panel (a) it is purely attractive 
($\delta=0, \Omega/2, \sqrt{2}\Omega$), while 
in (b) it is either purely gravitational 
($\delta=\sqrt{3}\Omega$) or expulsive ($\delta = 2\Omega$).
Bottom panels: Evolution of the center of the 
bright soliton for the above cases: (c) 
for attractive effective potentials and (d) 
for gravitational or expulsive ones. 
The agreement between numerical results 
(solid and dashed lines) and the theoretical predictions 
(triangles, dots) is excellent.
}
\label{fig1}
\end{figure}

In Fig. \ref{fig1}(a), the original harmonic trapping potential $V(x)$ (dotted line, $\delta=0$) is compared to 
the effective potential modified by the presence of the gradient for $\delta=(1/2)\Omega$ (dashed line) and 
$\delta=\sqrt{2} \Omega$ (solid line). In these cases, the effective harmonic potential is attractive and the 
gradient displaces the equilibrium point to the right. Fig. \ref{fig1}(b) shows the case 
$\delta=\sqrt{3}\Omega$ (dashed line) for which the effective harmonic potential is canceled resulting in a purely
gravitational potential.  Also, upon suitably choosing the value of 
$\delta$, e.g., for $\delta=2 \Omega$ (solid line) the effective potential becomes expulsive. The dynamics 
of the bright matter-wave soliton pertaining to the above cases are shown in Figs. \ref{fig1}(c) (for attractive effective 
potential) and \ref{fig1}(d) (for gravitational or expulsive effective potential). In particular, in Fig. \ref{fig1}(c), 
it is clearly seen that the evolution of the soliton center $x_0$ is periodic, but with a larger amplitude and smaller frequency of 
oscillations, as compared to the respective case with $\delta=0$. The analytical predictions of 
Eq. (\ref{eq_mot})-(\ref{soln}) 
(triangles for $\delta=\sqrt{2}\Omega$ and 
dots for $\delta=(1/2)\Omega$) are in perfect agreement with the 
respective results obtained by direct numerical integration of the GP Eq. (\ref{GP}). On the other hand, as shown in Fig. \ref{fig1}(d),
in the case of a gravitational or expulsive effective potential, the function $x_{0}$(t) is monotonically increasing, with 
the analytical predictions being in excellent agreement with the numerical simulations.
For a purely gravitational or expulsive effective potential, Eq. (\ref{bright_par1}) shows that the amplitude (width) of 
the soliton increases (decreases) monotonically as well, which recovers the predictions of Ref. \cite{fka}. 
This type of evolution suggests that the bright soliton is compressed adiabatically in the presence of the gradient. 

\begin{figure}[tbp]
\includegraphics[width=10cm]{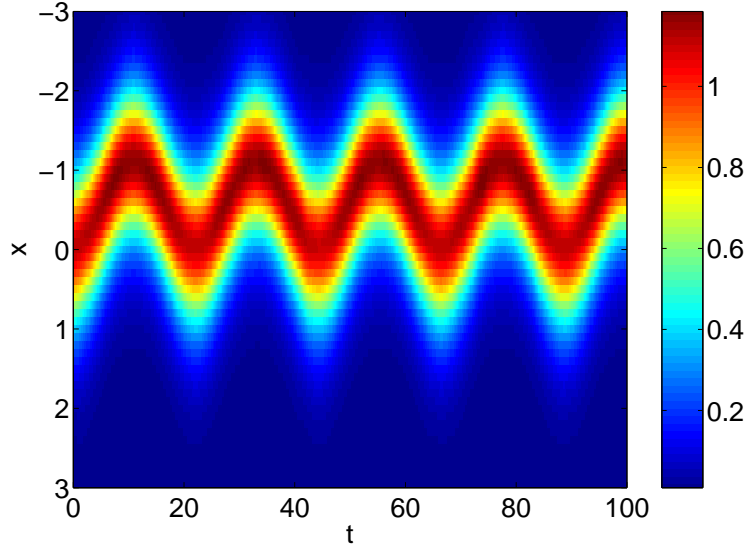}
\caption{Spatio-temporal contour plot (of the wavefunction
square modulus) of a solitary wave
for $\Omega=0.075$, $\delta=\sqrt{3} \Omega$ and an optical
lattice with $V_0=0.25$ and $\kappa =0.5$. One can clearly
discern the presence of Bloch oscillations in the evolution
of the density, whose period is in very good agreement with
the corresponding theoretical prediction.}
\label{pkfig2}
\end{figure}

Let us consider another setup which combines the ``effective'' linear potential 
with an external harmonic and a periodic trap:
\begin{eqnarray}
V(x)=\frac{1}{2} \Omega^2 x^2 + V_0 \sin^2 (\kappa x)
\label{ol}
\end{eqnarray}
The periodic potential in Eq. (\ref{ol}) 
can be obtained experimentally by superimposing
two counter-propagating laser beams. 
It is well-known that the dynamics in the combined
presence of a(n effective) linear and a periodic potential
results in the so-called Bloch oscillations
(for a recent discussion of the relevant phenomenology and bibliography
see e.g. \cite{bloch}). These oscillations occur due to 
interplay of the linear and periodic potential with a definite period
$T=2 \kappa/\beta$ \cite{bloch}.
We have examined numerically this analytical prediction in the
presence of an optical lattice potential with $V_0=0.25$ and 
$k=0.5$. The numerical evaluation of the period of the soliton
motion in the combined potential is $T \approx 22.15$ less than 
$4 \%$ off the corresponding theoretical prediction.
The time-periodic evolution of the soliton is shown in 
the spatio-temporal contour plot of Fig. \ref{pkfig2}.

\subsection{Higher-order solitons}

Apart from the fundamental bright soliton in Eq. (\ref{ansatz}), it is well known \cite{sy} that specific $\cal{N}$-soliton exact
solutions in the unperturbed NLS equation (Eq. (\ref{gpe1d_u}) with $s=-1$ and $R=0$) are generated by the initial condition
$u(x,0)= A {\rm sech}(x)$ (for $\eta=1$), and the soliton amplitude $A$ is such that $A-1/2 < \cal{N}$$\le A+1/2$ to excite 
a soliton of order $\cal{N}$. The exact form of the $\cal{N}$-soliton is cumbersome and will
not be provided here; nevertheless, it is worth 
noticing some features of these solutions: 
First, the number of atoms of the 
$\cal{N}$-th soliton is ${\cal N}^{2}$ larger than the 
one of the fundamental soliton and second, for any $\cal{N}$, the soliton solution is periodic with the intrinsic frequency of 
the shape oscillations being $\omega _{{\rm intr}}=4\eta^{2}$. 
We now examine  the dynamics of the $\cal{N}$-soliton solution in 
the presence of the spatially varying nonlinearity.

\begin{figure}[tbp]
\includegraphics[width=9cm]{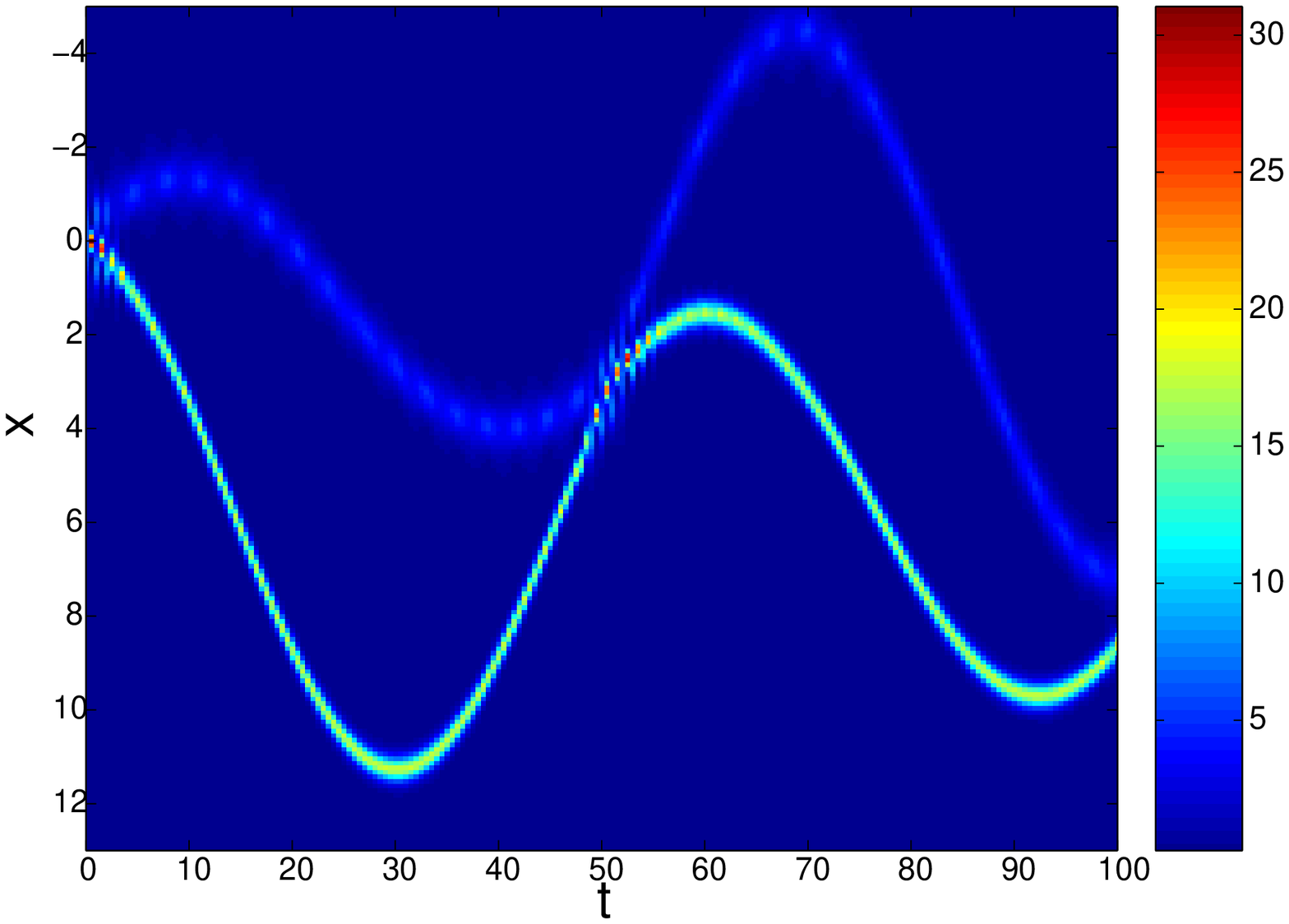}
\includegraphics[width=9cm]{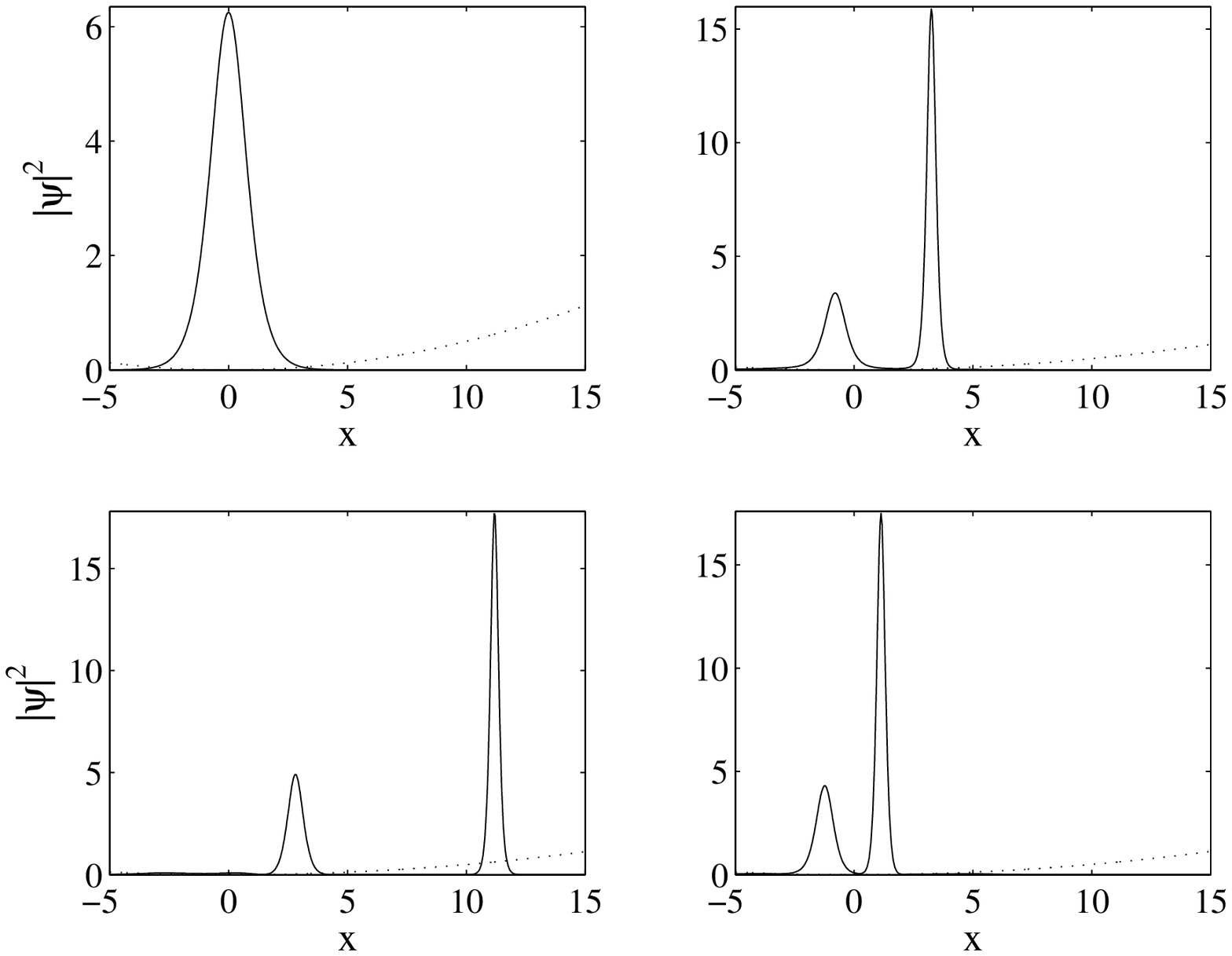}
\caption{Evolution of a double soliton with initial amplitude $A=2.5$ initially placed at the trap center ($x=0$) 
with a strength $\Omega=0.1$ in the presence of a gradient $\delta=0.01$. 
Top panel: Spatio-temporal contour plot of the 
density. Bottom panels: Snapshots of the evolution of the density 
are shown for $t=0$ (top left panel), $t=10$ (top right panel), $t=30$ (bottom left panel), and 
$t=60$ (bottom right panel), covering almost one period of the oscillation. Dashed lines correspond to the trapping potential.}
\label{fig1a}
\end{figure}

We have performed numerical simulations in the case of the so-called double ($\cal{N}$$=2$) 
bright soliton solution with initial soliton amplitude $A=2.5$. In the absence of the gradient ($\delta=0$), 
if the soliton is placed at the trap center ($x_{0}=0$, 
with $x_{0}$ being the soliton center), it only executes its 
intrinsic oscillations with the above mentioned frequency 
$\omega_{{\rm intr}}$. On the other hand, if the soliton is displaced 
($x_{0} \ne 0$), apart from its internal vibrations, 
it performs oscillations governed by the simple equation 
$\ddot{x}_{0}+\Omega^{2} x_{0}=0$, in accordance to the Kohn theorem 
(see \cite{kohn} and \cite{alk} for an application 
in the context of bright matter-wave solitons). Nevertheless, 
for $\delta \ne 0$, the double soliton (initially placed at the 
trap center), contrary to the previous case, splits into two single 
solitons, with different amplitudes due to 
the effective gravity discussed in the case of the fundamental soliton. Due to the effective gravitational force, 
the soliton moving to the right (see, e.g., Fig. \ref{fig1a})  
is the one with the larger amplitude (and velocity) and 
is more mobile than the one moving to the left (which has the smaller amplitude). 

As each of these two solitons is 
close to a fundamental one, their subsequent dynamics (after splitting) 
may be understood by means of the effective equations of motion derived in the previous section. In particular, depending on the 
values of the relevant parameters involved in Eq. (\ref{Veff}) [$\eta(0)$ is now the amplitude of each soliton after splitting] 
the solitons may both be trapped, or may escape (either one or both of them), 
if the effective potential is 
expulsive. In the former case, both solitons perform oscillations (in the presence of the effective attractive potential) and an 
example is shown in Fig. \ref{fig1a} (for $\Omega=0.1$, $\delta=0.01$). Note that the center of mass of the ensemble 
oscillates with a period $T=2 \pi /\Omega=62.8$ (which is in accordance with Kohn's theorem). 
During the evolution, as each of the two solitons
oscillate in the trap with different frequencies, they may undergo a head-on collision (see, e.g., bottom panel of Fig. \ref{fig1a} at 
$t \approx 55$). It is clear that such a collision is 
nearly elastic, with the interaction between the two solitons being repulsive. 

Importantly, for smaller values of the trap strength $\Omega$, we have found that it is possible to release either 
one or both solitons from the trap: In particular, 
for $\Omega=0.05$ and $\delta=0.025$, we have found that the large amplitude soliton escapes the trap, while the small-amplitude one 
performs oscillations. On the other hand, for the same value of the trap strength but 
for $\delta=0.05$  both solitons experience an expulsive effective potential and thus 
both escape from the trap. It is therefore in principle possible to use 
the spatially varying nonlinearity
not only to split a higher-order bright soliton to a chain of fundamental ones, but also to control the 
trapping or escape of the resulting individual fundamental solitons. 

\section{Dark matter-wave solitons}

We now turn to the dynamics of dark matter-wave solitons in the 
framework of Eq. (\ref{GP}) for $s=+1$ (i.e., the defocusing case of
condensates with repulsive interactions). Firstly we 
examine the equation governing the background wavefunction. 
The latter is taken in the form 
$\psi= \Phi(x) \exp(-i\mu t)$ ($\mu$ being the chemical potential) and the unknown background wave function 
$\Phi(x)$ satisfies the following real equation,
\begin{equation}
\mu\Phi+\frac{1}{2}\frac{d^{2} \Phi}{dx^2}-g(x) \Phi^{3}=V(x)\Phi.
\label{ub1}
\end{equation}
To describe the dynamics of a dark soliton on top of the inhomogeneous background satisfying Eq. (\ref{ub1}), 
we introduce the ansatz (see, e.g., \cite{d})
\begin{equation}
\psi=\Phi(x) \exp(-i\mu t)\upsilon (x,t) , 
\label{ansatz2}
\end{equation}
into Eq. (\ref{GP}), where the unknown wavefunction $\upsilon (x,t)$ represents a dark soliton. 
This way, employing Eq. (\ref{ub1}), the following evolution equation for the dark soliton wave function is readily obtained:
\begin{equation}
i\frac{\partial \upsilon}{\partial t} +\frac{1}{2} \frac{\partial^{2} \upsilon}{\partial x^2}
-g \Phi^{2}(|\upsilon|^{2}-1)\upsilon = -\frac{d}{dx}\ln(\Phi) \frac{\partial \upsilon}{\partial x}.
\label{ups}
\end{equation}
Taking into account that in the framework of the Thomas-Fermi approximation \cite{dalfovo} 
a simple solution of Eq. (\ref{ub1}) is expressed as 
\begin{equation}
\Phi(x)=\sqrt{\max\{\frac{\mu -V(x)}{g(x)},0\}},  
\label{TF}
\end{equation}
equation (\ref{ups}) can be simplified to the following  
defocusing perturbed NLS equation,
\begin{equation}
i\frac{\partial \upsilon}{\partial t}+\frac{1}{2}\frac{\partial^{2} \upsilon}{\partial x^2}-\mu(|\upsilon |^{2}-1)\upsilon=Q(\upsilon),
\label{pnls}
\end{equation}
where the perturbation $Q(\upsilon)$ has the form, 
\begin{eqnarray}
Q(\upsilon)&=& \left( 1-|\upsilon |^{2}\right) \upsilon V+ \frac{1}{2(\mu-V)}\frac{dV}{dx} \frac{\partial \upsilon}{\partial x} 
\nonumber \\
&+& \frac{d}{dx}\left[ \ln(\sqrt{g})\right] 
\frac{\partial \upsilon}{\partial x},
\label{pert}
\end{eqnarray}
and higher order perturbation terms have once again been neglected.
In the absence of the perturbation, Eq. (\ref{pnls}) represents the completely integrable defocusing NLS 
equation, which has a dark soliton solution of the form \cite{zsd} (for $\mu=1$),
\begin{eqnarray}
\upsilon (x,t)=\cos \varphi \tanh \zeta +i \sin \varphi,
\label{ds}
\end{eqnarray}
where $\zeta \equiv \cos \varphi \left[ x-(\sin \varphi)t \right]$, while $\cos \varphi$ and $\sin \varphi$ are the 
soliton amplitude and velocity respectively, $\varphi$ being the so-called soliton phase angle ($|\phi| \le \pi/2$). 
To treat analytically the effect of the perturbation (\ref{pert}) on the dark soliton, we employ the adiabatic 
perturbation theory devised in Ref. \cite{KY}. As in the case of bright solitons, according to this approach, 
the dark soliton parameters become slowly-varying unknown functions of $t$, but the functional form of the soliton 
remains unchanged. Thus, the soliton phase angle becomes $\varphi \rightarrow\varphi(t)$ and, as a result, 
the soliton coordinate becomes 
$\zeta \rightarrow \zeta=\cos\varphi(t) \left[x-x_{0}(t) \right]$, where
\begin{equation}
x_{0}(t)= \int_{0}^{t}\sin\varphi(t^{\prime })dt^{\prime},
\label{cent}
\end{equation}
is the soliton center. Then, the evolution of the parameter $\varphi$ governed by the equation \cite{KY},
\begin{equation}
\frac{d\varphi}{dt}=\frac{1}{2\cos ^{2}\varphi \sin \varphi} {\rm Re}
\left[ \int_{-\infty}^{+\infty}Q(\upsilon)\frac{\partial \upsilon^{\ast}}{\partial t} dx \right],
\label{phi}
\end{equation}
leads (through similar calculations and Taylor expansions as for the 
bright case) to the following result:
\begin{eqnarray} 
\frac{d \phi}{dt}=-\cos\varphi \left[ \frac{1}{2} \frac{\partial V}{\partial x_{0}}+\frac{1}{3} 
\frac{\partial}{\partial x_{0}} \ln(g) \right].
\label{dark_par2} 
\end{eqnarray} 
To this end, combining Eqs. (\ref{cent}) and (\ref{dark_par2}), we obtain the
corresponding equation of motion for the 
soliton center,
\begin{eqnarray} 
\frac{d^2x_{0}}{dt^2}=-\frac{1}{2} \frac{\partial V}{\partial x_{0}}
-\frac{1}{3} \frac{\partial}{\partial x_{0}} \ln(g), 
\label{eqm} 
\end{eqnarray} 
in which we have additionally assumed nearly stationary dark solitons with $\cos \varphi \approx 1$. 
As in the case of bright solitons, the validity of Eq. (\ref{eqm}) does not rely on the specific form of $g(x)$, 
as long as this function (and the trapping potential) are slowly-varying on the dark soliton scale (i.e., the healing length). 
In the particular case with $g(x)=1+\delta x$, Eq. (\ref{eqm}) describes the motion of a unit mass particle 
in the presence of the effective potential  
\begin{eqnarray} 
W_{\rm eff}(x_{0})=\frac{1}{4} \Omega^{2} x_{0}^{2}+\frac{1}{3} \ln(1+\delta x_{0}).
\label{Vef} 
\end{eqnarray}
For $\delta=0$ Eq. (\ref{eqm}) implies that the dark soliton oscillates with a frequency $\Omega/\sqrt{2}$ 
in the harmonic potential with strength $\Omega$ \cite{d,motion}.  However, in the presence of the gradient, and for 
sufficiently small $\delta$, Eq. (\ref{Vef}) implies the following: First, the oscillation 
frequency $\omega_{\rm ds}$ of the dark 
soliton is downshifted in the presence of the linear spatial variation
of the scattering length, according to 
\begin{equation}
\omega_{\rm ds} = \sqrt{\frac{1}{2}\Omega^{2}-\frac{1}{3}\delta^{2}}.
\label{osd}
\end{equation} 
Additionally to the effective harmonic potential, 
the dark soliton dynamics is also modified by an effective 
gravitational potential ($\sim \delta x_0/3$), which  
induces an acceleration of the soliton towards larger values of $x_0$ (for $\delta >0$). It should be noted that as dark solitons 
behave as effective particles with negative mass, the effective gravitational force possesses a positive sign, while in the case of 
bright solitons (which have positive effective mass) it has the usual negative sign [see Eqs. (\ref{Veff}) and (\ref{Vef})].

\begin{figure}[tbp]
\includegraphics[width=6cm]{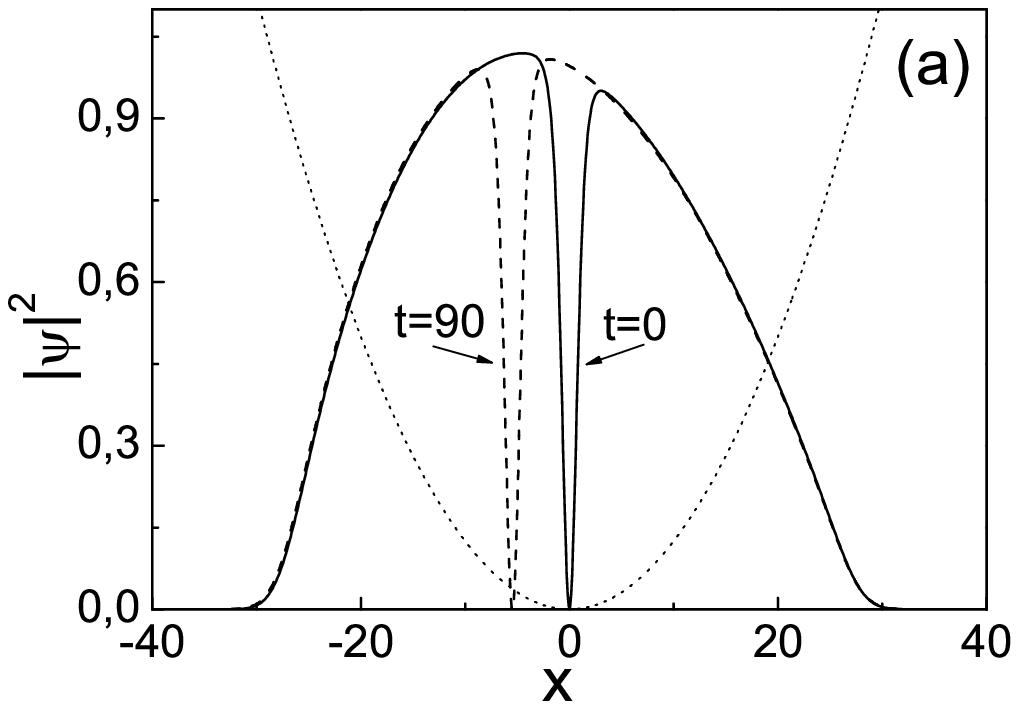}
\includegraphics[width=6cm]{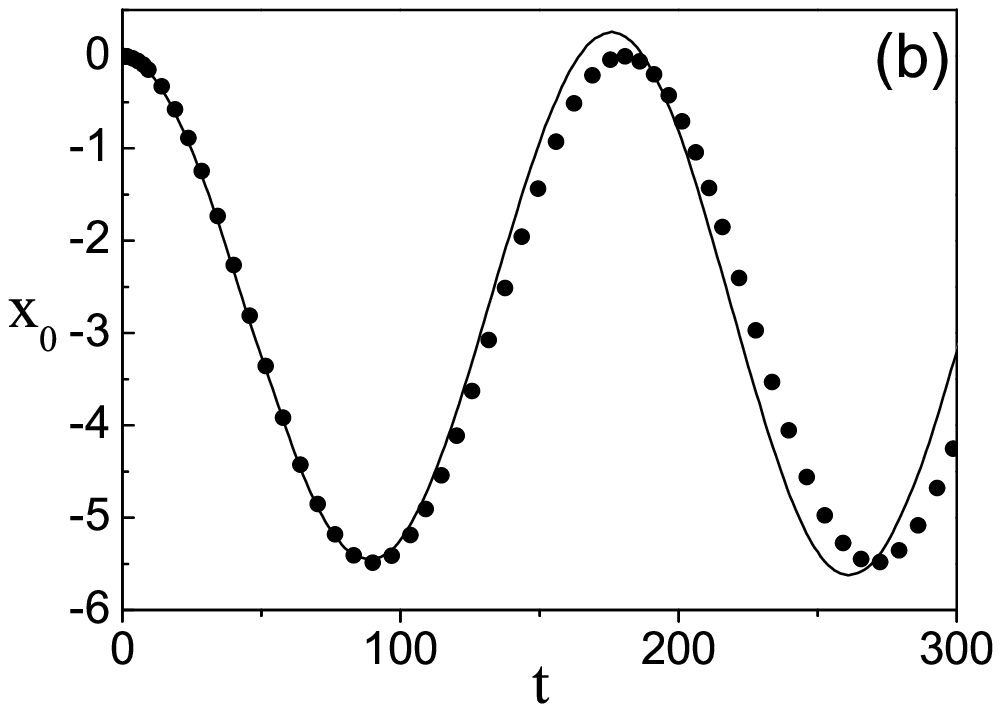}
\caption{
(a) Two snapshots of the density of the dark soliton (at $t=0$ and $t=90$) on top of a Thomas-Fermi cloud.
The chemical potential is $\mu=1$, the trap strength is $\Omega=0.05$ and the gradient is $\delta=0.01$.
(b): The motion of the center of a dark soliton. Solid line and dots respectively correspond to the numerical integration 
of the GP equation and analytical predictions [see Eq. (\ref{eqm})], respectively.  
}
\label{fig2}
\end{figure}

Direct numerical simulations  confirm the 
above analytical findings. In particular, we consider an initially  
stationary dark soliton (with $\cos \varphi(0)=0$), placed at $x_{0}=0$, on top of a Thomas-Fermi cloud [see Eq. (\ref{TF})] 
characterized by a chemical potential $\mu=1$ (the trapping frequency is here $\Omega=0.05$). In the absence of 
the gradient such an initial dark soliton should be purely stationary. However, considering a 
gradient with $\delta=0.01$, it is clear that the TF cloud will become asymmetric, as shown in Fig. \ref{fig2}(a) and 
the soliton will start performing oscillations. The latter are shown in Fig. \ref{fig2}, where the analytical predictions (points) 
are directly compared to the results obtained by direct numerical integration of the GP equation (solid line). As it is seen,
the agreement between the two is very good; additionally, we note that the oscillation frequency found numerically is 
$2 \pi/177$, while the respective theoretical prediction is $2 \pi/180.7$, with the error being $\approx 3\%$.

\section{Summary}

We have analyzed the dynamics of bright and dark matter-wave solitons in quasi-1D BECs characterized by  
a spatially varying nonlinearity. The formulation of the problem is based on a Gross-Pitaevskii equation with a spatially dependent 
scattering length induced e.g. by a bias magnetic field near a Feshbach resonance augmented by a
field gradient.
The GP equation has been reduced to a perturbed nonlinear Schr{\"o}dinger equation, which is then 
analyzed in the framework of the adiabatic approximation 
in the perturbation theory for solitons, treating them as 
quasi-particles. This way, effective equations of motion for the soliton centers (together with evolution equations for their other  
characteristics) were derived analytically. The analytical results were corroborated by direct numerical simulations of the 
underlying GP equations.

In the case of bright matter-wave solitons initially confined in a parabolic trapping potential, it is found that  
(depending on the values of the gradient and the initial soliton parameters), there is a possibility 
to switch the character of the effective potential from attractive to purely gravitational or expulsive. It has been thus 
demonstrated that a bright soliton can escape the trap and be adiabatically compressed. On the other hand, considering the 
additional presence of an optical lattice potential, it has been shown
that in the case where the effective potential is purely gravitational,
Bloch oscillations of the bright solitons are possible. Higher-order bright solitons have been shown to typically split
in the presence of a spatially varying nonlinearity to fundamental ones, whose subsequent 
dynamics is determined by the properties of the resulting single-soliton splinters. 
In the case of dark matter-wave solitons, the relevant background, 
i.e., the Thomas-Fermi cloud is modified by the inhomogeneous nonlinearity. The dynamics of the dark solitons 
follows a Newtonian equation of motion for a particle with a negative effective mass and the 
oscillation frequency of the dark solitons has been derived analytically. The latter is always down-shifted 
as compared to the oscillation frequency pertaining to a spatially constant scattering length. 
Thus, generally speaking, the presented results show that a spatial inhomogenity of the scattering length induced e.g. by properly 
chosen external magnetic fields is an effective way to control the dynamics of matter-wave solitons.

{\bf Acknowledgements.} This work was supported by 
the ``A.S. Onasis'' Public Benefit Foundation (GT),
the Special Research Account of Athens University (GT, DJF), 
as well as NSF-DMS-0204585, NSF-CAREER, and the Eppley Foundation for Research (PGK).


\end{document}